\documentclass[aps,twocolumn]{revtex4}
\usepackage{mathrsfs}
\usepackage{amsmath}
\usepackage{graphicx}
\usepackage{subfigure}
\usepackage{epstopdf}

\begin{document}

\renewcommand{\bottomfraction}{.7}
\setcounter{topnumber}{12} \setcounter{bottomnumber}{12}
\setcounter{totalnumber}{20} \setcounter{dbltopnumber}{12}





\title{The chiral condensate in a constant electromagnetic field}

\author{Thomas D. Cohen}
\email{cohen@physics.umd.edu}

\author{David A. McGady}
\email{dmcgady@umd.edu}

\author{Elizabeth S. Werbos}
\email{ewerbos@physics.umd.edu}

\affiliation{Department of Physics, University of Maryland,
College Park, MD 20742-4111}

\begin{abstract}
We study the shift of the chiral condensate in a constant
electromagnetic field in the context of chiral perturbation
theory. Using the Schwinger proper-time formalism, we derive a
one-loop expression correct to all orders in $m_{\pi}^2 / eH$. Our
result correctly reproduces a previously derived ``low-energy
theorem'' for $m_\pi = 0$. We show that it is essential to include
corrections due to non-vanishing $m_\pi$ in order for a low energy
theorem to have any approximate regime of validity in the physical
universe. We generalize these results to systems containing
electric fields, and discuss the regime of validity for the
results. In particular, we discuss the circumstances in which the
method formally breaks down due to pair creation in an electric
field.

\end{abstract}

\maketitle

\section{Introduction}

Quantum Chromodynamics (QCD) provides the basis for our
understanding of hadronic matter at its most fundamental level. As
such, mathematically stable solutions to the exact equations of
QCD represent stable physical particles.  Because extreme external
conditions such as high temperatures, strong chemical potentials,
or powerful electromagnetic fields couple to quark and gluon
fields, they alter the mathematically stable solutions of QCD.
Calculating the corresponding response of the various physical
observables of hadronic matter in the presence of these extreme
conditions is a central question at the core of modern nuclear
physics.  Research along these lines has mainly focused on the
effects of high temperature or high density (or equivalently
chemical potential) on QCD matter.  The response of QCD
observables to very large electromagnetic fields has undergone
less extensive study, but has potentially equally interesting
consequences.

Since electromagnetic fields couple directly to quarks and not to
gluons, it is natural to focus on intensive observables built from
quark fields. It is clearly essential to understand the behavior
of the chiral condensate, $\Sigma = \langle \overline{q}q
\rangle$, since the chiral condensate quantifies one of the most
salient features of QCD---the spontaneous breaking of chiral
invariance. To date the response of the chiral condensate to
either constant electric and constant magnetic fields has only
been studied in a variety of low energy models and effective field
theories. All calculations of the shift of the chiral condensate
in the presence of strong magnetic fields yield an enhanced chiral
condensate. These calculations include those done in the various
low energy models of QCD at or close to the chiral limit
\cite{Gorbar, Goyal, Klevansky, NJL,SmilgaFurther,Gusynin} and in the model-independent
calculations \cite{Smilga,Shushpanov} of chiral
perturbation theory ($\chi$PT) \cite{GasserLeutwyler}. Shifts due
to ``constant'' electric fields are not as well studied. Even
though calculations made in the Nambu-Jona-Lasinio (NJL) model
\cite{NJL,Klevansky,Gorbar,SmilgaFurther,Gusynin} show that a constant electric field
reduces the chiral condensate, it seems that this has yet to be
verified in a model-independent calculation.

All of the calculations generally have an enhanced spontaneous
chiral symmetry breaking (S$\chi$SB) due to a magnetic field. We
choose to focus here on $\chi$PT, following \cite{Smilga}, because
as the effective theory for QCD its results should be more
generally applicable. Shushpanov and Smilga\cite{Smilga} find a
result which they call a ``low-energy theorem,'' holding in the
strict chiral limit at one loop in $\chi$PT, to be
\begin{equation}
\label{eqn:lowenergy} \frac{\Sigma(H)}{\Sigma(0)} =  1+
\frac{\log(2) \, eH}{16 \pi^2 F_{\pi}^2}.
\end{equation}
By design, $\chi$PT is {\it universally} applicable to all
theories with underlying chiral symmetry and spontaneous chiral
symmetry breaking.

Thus for fields sufficiently small, the $\chi$PT result $\propto
eH$ will be the leading order result. The principal results of
this paper is the generalization of the ``low energy
theorem''\cite{Smilga} in Eq.~(\ref{eqn:lowenergy}). This result
has been extended to $\chi$PT at two loops \cite{Shushpanov},
increasing the accuracy of the theoretical predictions for large,
pure magnetic fields in the limit where $m_\pi=0$. It is important
to note that the assumption that $m_\pi = 0$ may place a severe
constraint on these results. While results obtained in this
theoretical limit are certainly interesting, it is not necessary
to impose such a condition in order to formulate a controlled
expansion. Consistent chiral expansions can only be formulated
when both $e H$ and $m_\pi^2$ can be treated as low mass scales
compared to $\Lambda^2$.  However, nothing in the formulation
fixes the ratio $m_\pi^2/eH$ which can be kept arbitrary. The low
energy theorem of Eq.~(\ref{eqn:lowenergy}) must be regarded as
the leading term in the full chiral expansion restricted to a
regime where $e H \gg m_\pi^2$.

Physically, $m_\pi \sim 140$ MeV, and it is an open question
whether the result derived at $m_\pi=0$ accurately describes
observable shifts in nature for any given values of $H$. Indeed,
it is unclear that there is {\it any} domain of validity for
Eq.~(\ref{eqn:lowenergy}), since it is not clear that it is
possible to simultaneously have $e H \gg m_\pi^2$ while allowing
$e H$ to remain within the regime where chiral perturbation theory
to one loop is accurate.  Most simply, the condition for $\chi$PT
to be applicable is $p^2/\Lambda^2 \ll 1$, where $p$ is any
relevant light scale in the problem--mass, magnetic field,
external momenta, etc. In practice, one might generically expect
$m_\pi^2/\Lambda^2 \sim 1/50$, which is a small expansion
parameter when only one condition is required. Difficulties arise,
however, when one artificially imposes the extra condition that
$m_\pi^2/eH \ll 1$, which will then in practice require the tight
hierarchy $m_\pi^2/\Lambda^2 \sim 1/50 \ll eH/\Lambda^2 \ll 1$.
The region of magnetic fields which satisfy this hierarchy will be
at best rather narrow.

We will show by explicit calculation, the general low energy
theorem, valid for all values of $m_\pi^2/eH$, converges quite
slowly to Eq.~(\ref{eqn:lowenergy}). It seems apparent that if an
$H$ field is large enough to be in the regime of validity of
Eq.~(\ref{eqn:lowenergy}), it would also be sufficiently large as
to require the inclusion of higher-order operators in the chiral
Lagrangian. As a practical matter it is certainly far more useful
to obtain a result valid for $m_\pi^2 \ll \Lambda^2, e H \ll
\Lambda^2$, and arbitrary $m_\pi^2/eH$. Such a generalization
greatly extends the regime of validity. Thus, our principal result
of providing a ``low energy theorem'' valid at all orders is
$m_\pi^2/eH$ is essential for having any useable result for the
physical world.  Moreover, one expects that the coefficients are
likely to be rather unfavorable in this case, given $1/N_c$
considerations which we will discuss briefly in the conclusion.

\begin{figure}[htbp]
  \includegraphics[width=0.99\linewidth]{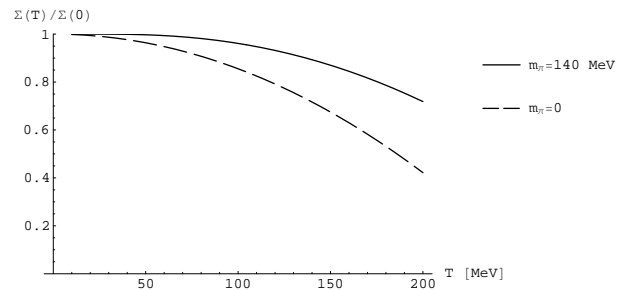}
\caption{\label{fig:Tplot}Shift in the condensate to one loop
plotted as a function of temperature in the chiral limit and with
a realistic finite value for $m_\pi$.}
\end{figure}

An analogous situation can be found in the example of ``low-energy
theorems'' for QCD at finite temperature. We present it here to
emphasize that the theoretically elegant limit of $m_\pi = 0$ is
not always relevant in practice. Ref.~\cite{Smilga} argues that
Eq.~(\ref{eqn:lowenergy}) is a theorem in exactly the same sense
that
\begin{equation}
\frac{\Sigma(T)}{\Sigma(0)} = 1- \frac{T^2}{8 F_{\pi}^2}
-\frac{T^4}{384 F_\pi^4} - ... \label{eqn:T}
\end{equation}
is a low energy theorem for the condensate which holds at low $T$
in the strict chiral limit \cite{GasserLeutwylerT,Gerber}. Clearly
Eq.~(\ref{eqn:T}) is formally valid only when the hierarchy of
scales, $m_\pi \ll T \ll T_c \lesssim \Lambda_{QCD}$, is
satisfied.  The range of validity for this hierarchy is similar to
the one we find in our problem. Also in analogy, while
Eq.~(\ref{eqn:T}) is formally correct in the $m_\pi=0$ limit, it
is never useful in describing a real system at any temperature:
any temperature which is high enough to be much bigger than
$m_\pi$ is also beyond the temperature of the QCD phase transition
and thus outside the regime of validity of $\chi$PT. Since the
formula is derived with the assumption that $m_\pi = 0$ and
$m_\pi$ does not appear in the result, it must be the case that
the expression is only valid for $m_\pi \ll T$; if it is of the
same order as $T$ or greater, it will begin to play an
increasingly important role in the result. Thus, for $T$ of the
order of tens of MeV where (\ref{eqn:T}) would be valid, $m_\pi$
is comparatively large enough to render (\ref{eqn:T}) invalid.
Further, since $T_c \sim 170$ MeV, if $T \gg m_\pi \sim 140$ MeV
then one is clearly in the quark-gluon plasma phase, and outside
the range of $\chi PT$. Explicit calculations with the physical
value of $m_\pi$ \cite{Gerber} show conclusively that the``low
energy theorem'' of Eq.~(\ref{eqn:T}) does not accurately
reproduce the shift in the chiral condensate for any temperature.
We illustrate this explicitly in Fig.~\ref{fig:Tplot}.

The approach to implementing $\chi$PT at one-loop (${\cal
O}(p^4)$) for our problem is straightforward.  At ${\cal O}(p^4)$
the pions do not interact, and one uses the appropriate
non-interacting propagators for a constant external field. In this
circumstance, the Schwinger proper time formalism \cite{Schwinger}
provides a natural framework to study the QCD chiral condensate in
the presence of constant electromagnetic fields.

The final issue we address in this paper is the generalization of
the low energy theorem to include electric fields---either as pure
electric fields or in situations where both $E$ and $H$ are present.
The character of the shift of the condensate in the presence of a
non-trivial electric field is fundamentally different from the shift
in the presence of a pure magnetic field. This difference is
essentially due to the famous Schwinger mechanism, and is made
manifestly clear within the context of the proper time formalism.
Within this formalism, it is a relatively straightforward exercise
to evaluate the effective action for charged matter fields in the
presence of a uniform electromagnetic field (to one loop). Poles
appear in the effective action in the presence of a uniform electric
field, which are conventionally interpreted as corresponding to
spontaneous real $\pi^+\pi^-$ pair creation out of the vacuum
\cite{Schwinger}. This in turn implies an inherent local instability
in a system containing a constant electric field. Indeed, in light
of the Schwinger mechanism, a uniform field in such a system does
not remain static, but naturally evolves via back-reactions over
time \cite{CooperMottola,Smolyansky}. As a practical matter, the
effect of pair creation in a constant electric field means that the
field can only be considered as constant over length and time scales
limited by the parameters of the problem. In contrast, a system with
a constant magnetic field has no such instability, and the $H$ field
can consistently be regarded as constant over time.

\section{The chiral condensate in an external magnetic field \label{sec:mag}}

In this section we compute the chiral condensate in an external
magnetic field to one loop in chiral perturbation theory and to
all orders in $m_\pi^2/e H$.  Recall that the chiral expansion is
typically an expansion in $\frac{m_\pi^2, p^2}{\Lambda^2}$, where
$p$ is a momentum in the problem.  Since the external magnetic
field is an isovector, it explicitly breaks chiral symmetry, and
the expansion becomes an expansion in  $\frac{m_\pi^2, p^2, e
H}{\Lambda^2}$. Since the scattering amplitude for pions at
$p^2=0, m_\pi^2=0, H=0$ is zero at lowest order in the theory, the
leading-order $\chi$PT result for the shift in the chiral
condensate due to a magnetic field is simply the one-loop
expression for non-interacting pions in a magnetic field.  In
conventional $\chi$PT counting, one-loop effects occur at ${\cal
O}(p^4)$. The effects of pion-pion interactions occur at higher
order in the chiral expansion and will accordingly be suppressed
by powers of $\frac{m_\pi^2,e H}{\Lambda^2}$.

The exact expression \cite{Schwinger} for the effective Lagrangian
for a charged pion in an external field is given by
\begin{equation} \label{eqn:SchwingerLeffBoson}
{\cal L}_{eff} = \frac{1}{16\pi^2}\int_{0}^{\infty}ds s^{-3}
e^{-m_\pi^2 s}\left(\frac{(es)^2{\cal G}}{{\cal I}(\cosh esX)} -
1\right),
\end{equation}
 where ${\cal F}=
\frac{H^2-E^2}{2}$ and ${\cal G }=\vec{E} \cdot \vec{H}$ and $X =
({\cal F} + i {\cal G})^{\frac{1}{2}}$. In an $H$ field, this can
be simplified to
\begin{equation} \label{eqn:SchwingerMagneticBoson}
{\cal L}_{\rm eff} = \frac{1}{16\pi^2}\int_{0}^{\infty}
\frac{ds}{s^3}e^{-m_{\pi}^2 s} \left[ \frac{eHs}{\sinh(eHs)} - 1
\right].
\end{equation}
As noted above, this corresponds to the one-loop $\chi$PT result.
Thus, the one-loop chiral condensate as a function of applied
magnetic field is given by
\begin{equation}
\Delta \Sigma (H) =  \frac{\partial {\cal L}_{\rm eff}(H,
m_u)}{\partial m_u}.
\end{equation}

At this order, the mass of the pion is related to the quark mass,
the chiral condensate at zero field, and the pion decay constant
via the Gell-Mann-Oakes-Renner relation: $(m_u+m_d)\Sigma(0) =
F_\pi^2m_\pi^2$.  Therefore, the expression for the shift in the
condensate can be expressed as \cite{Smilga}
\begin{equation} \label{eqn:MagInt}
\begin{split}
\Delta \Sigma (H) =  \frac{\partial {\cal L}_{\rm eff}}{\partial
m_u} =  \frac{\log (2)\, eH\Sigma(0)}{16 \pi^2 F_{\pi}^2} I_H
\left( \frac{m_{\pi}^2}{ e H} \right )\\
I_H (y)  \equiv -\frac{1}{\log (2)} \int_{0}^{\infty} \frac{dz}{z^2}
e^{-y z} \left[ \frac{z}{\sinh(z)} - 1 \right] \, ,
\end{split}\end{equation}
where the parameter $y$ is the dimensionless ratio $m_\pi^2/eH$. A
direct comparison with Eq.~(\ref{eqn:lowenergy}) makes the physical
meaning of $I_H \left( \frac{m_{\pi}^2}{ e H} \right )$ clear: it is a
multiplicative factor which encodes the corrections to
Eq.~(\ref{eqn:lowenergy}) due to a non-zero $m_\pi$ to all orders in
$m_\pi^2/eH$.  The form of this integral simplifies dramatically in
the chiral limit, where $y \to 0$.  After a routine calculation it is
easy to see that $I_H(0)$ is unity, reproducing the low energy
theorem originally found in Ref.~\cite{Smilga}. However, for our
purposes, it is more interesting to note that there is a closed-form
expression for this integral \cite{GR}:
\begin{widetext}
\begin{equation} \label{eqn:exactH}
I_H (y) =\frac{1}{\log 2}\left( \log(2 \pi) +y\log \left(\frac{y}{2}\right) - y
- 2 \log \Gamma\left(\frac{1 + y}{2}\right)\right),
\end{equation}
\end{widetext}
yielding an analytic result to one-loop order in $\chi$PT valid for
any ratio of $m_\pi^2$ to $eH$.

A comment about $\chi$PT to one-loop is useful at this stage.
Typically, one-loop graphs diverge and are only sensible in the
context of a renormalization scheme including counterterms in the
${\cal O}(p^4)$ Lagrangian. The present expression is finite. {\it
A priori} this does not mean that there cannot be a finite
contribution from a higher-order operator in the chiral
Lagrangian. Direct inspection of the terms in the chiral
Lagrangian at ${\cal O}(p^4)$ reveals that no terms contribute to
the shift in the condensate in an external field at tree level. As
it happens, the first such term which contributes to the chiral
condensate at tree level occurs at ${\cal O}(p^6)$. Accordingly,
the lowest-order chiral Lagrangian, ${\cal O}(p^2)$, is sufficient
to determine the shift in the condensate to the order which we are
working.

\begin{figure}[tbp]
  \includegraphics[width=0.99\linewidth]{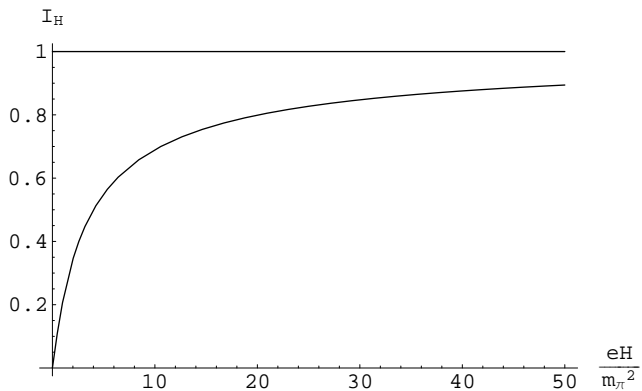}
\caption{\label{fig:Iplot}Exact expression for the integral
representing the shift in the condensate plotted as a function of
$eH/m_\pi^2$, as compared to the $m_\pi=0$ value of unity.}
\end{figure}

In Fig.~\ref{fig:Iplot}, $I_H(y)$ is plotted as a function of $1/y$.
It is clear from this figure that the convergence to unity is very
slow.  For example, even when $e H/m_\pi^2$ is 30, the exact
one-loop expression is only 85\% of the $m_\pi=0$ expression. This
value for $eH$ is already pushing the $eH/m_{\pi}^2 \ll
\Lambda^2/m_{\pi}^2 \sim 50$ scale mentioned earlier. We can
conclude, then, that any value of $eH$ large enough for the
``low-energy theorem'' of Ref.~\cite{Smilga} to be close to the
exact one-loop expression is likely to be near the region where the
small-field limit breaks down and higher-order terms in $\chi$PT are
important. Therefore, while the low-energy theorem is formally
correct, its assumptions remove its region of validity from the real
world, which has a finite $m_\pi$.  More importantly, the
generalized low energy theorem of Eqs.~(\ref{eqn:MagInt}) and
(\ref{eqn:exactH}) {\it is} valid over the full regime of small
fields and pion masses.

\section{The chiral condensate in an external electric field\label{sec:Efield}}

As noted in the Introduction, the $\chi$PT analysis to date has
been restricted to constant magnetic fields.  In contrast, the NJL
model calculations have been made for constant electric fields
\cite{Klevansky}.  Formally, these NJL calculations neglect an
imaginary part which arose in the evaluation of the chiral
condensate.  The standard interpretation of the emergence of an
imaginary part in the calculation of a purely real quantity is  a
signal that the state is not the true ground state of the system;
the state is regarded as unstable. In the mean-field NJL case the
instability is due to pair creation of constituent quarks. Since
this is a manifestation of the model's unphysical lack of
confinement, one may argue that this may be neglected. As we shall
see below an analogous issue arises in $\chi$PT at one loop, but
in this case the imaginary parts are due to an instability in
associated pion pair production and are undoubtedly physical.

Suppose that an electric field which is approximately uniform over
a large region of space is suddenly turned on. This external
electric field will cause both a real shift in the condensate and
a local breakdown in the vacuum due to real $\pi^+ \pi^-$ pair
emission. Provided that the instability due to pair emission
occurs over a much shorter time scale than the characteristic time
with which the condensate responds, it is sensible to discuss
shifts in the chiral condensate in the context of a ``constant''
electric field. Moreover, for small $E$ fields, the rate of pair
production scales with $m_\pi$ as $\exp \left ( -\frac{\pi
m_\pi^2}{e E} \right )$, and is exponentially suppressed
\cite{Schwinger}. Thus, in the regime where $e E \ll m_\pi^2$ time
evolution due to the imaginary shift is very slow and the question
of how the condensate responds to a spatially and temporally {\it
constant} electric field remains sensible.

The ratio of the calculated real to imaginary parts of the shift in
the condensate is a crude indicator of whether the calculation of the
real part is meaningful; we trust the result only if the real part
dominates.  When the imaginary part becomes comparable to the real
part, the instability is significant and the calculation of the real
part of the shift is unreliable.

Our goal in this section is to calculate the shift of the condensate
in an external electric field to one loop in $\chi$PT.  Qualitatively,
the electric field case and the magnetic field case are quite
different, though they are both derived from
Eq.~(\ref{eqn:SchwingerLeffBoson}). Because the chiral condensate is a
Lorentz scalar, shifts in the chiral condensate due to electromagnetic
fields can only depend on the Lorentz scalars ${\cal F}$ and ${\cal
G}$. Since ${\cal G} = 0$ for both constant electric and magnetic
fields, the shift in $\Sigma$ is only a function of $\cal{F}$. Due to
the exclusive dependence on ${\cal F}$, and because it is negative for
a constant $E$ field and positive for a constant $B$ field, one can
obtain the expression for a constant $E$ field by analytically
continuing the expression in Eqs.~(\ref{eqn:MagInt}),
(\ref{eqn:exactH}) from positive ${\cal F}$ to negative ${\cal
F}$. This analytic continuation is, in effect, the substitution $H
\rightarrow i E$.

This substitution $H \rightarrow i E$ induces the change in the
integrand $1/\sinh(z) \rightarrow 1/\sin(z)$. As such, the
integral acquires an infinite number of poles along the
integration path. Thus, we can write the shift as
\begin{equation}\label{eqn:ElecInt}
\begin{split}
\Delta \Sigma (E)  &= \frac{\log (2) \, eE\Sigma(0)}{16 \pi^2
F_{\pi}^2} I_E \left( \frac{m_{\pi}^2}{ e E} \right )
\\I_E(y)&\equiv -\frac{1}{\log (2) }\int_{0}^{\infty} \frac{dz}{z^2}
e^{-y z} \left[ \frac{z}{\sin(z)} - 1 \right] = - i I_H(i
y).
\end{split}
\end{equation}
The question of how to handle the poles in this integral is
intimately related to the boundary conditions imposed on the problem
which in turn necessarily reflect the underlying physical
circumstances.  Here, we will adopt the usual convention for the
Schwinger mechanism \cite{Schwinger}: we make the substitution
$1/\sin(z) \rightarrow 1/(\sin(z)+i\epsilon)$ in the integrand.
This renders the expression for the shift mathematically well
defined; physically it is the regime associated with pair creation
from the electric field.

The integral expression for $I_E$ makes it manifestly clear how
instabilities in the system arise in the presence of a uniform
electric field. The infinite number of poles with non-trivial
residues along the integration path allow the integral to be
separated into a real principal value part and a purely imaginary
part proportional to the sum of residues of the poles. Physically,
these poles, and the associated imaginary shift, indicate an
instability of the configuration and ultimately non-trivial
evolution of the system over time.

\begin{figure}[tbp]
\subfigure[$Re\{I_{E}\}$ vs $eE/m_\pi^2$]{
  \includegraphics[width=0.99\linewidth]{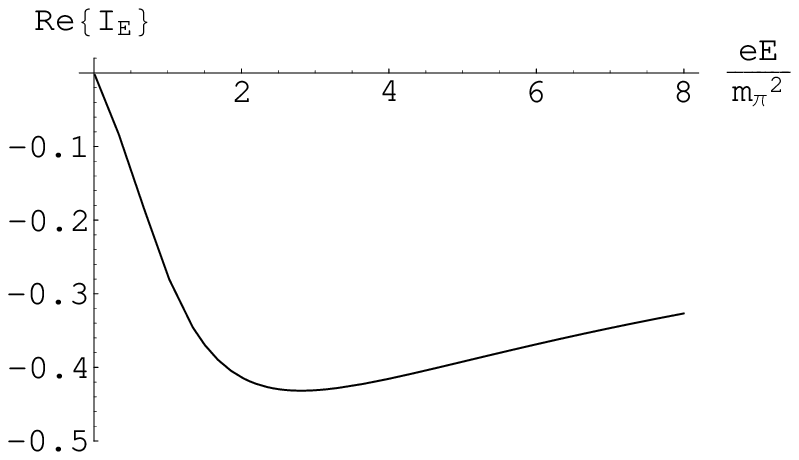}}
\\\subfigure[$Im\{I_{E}\}$ vs $eE/m_\pi^2$]{
  \includegraphics[width=0.99\linewidth]{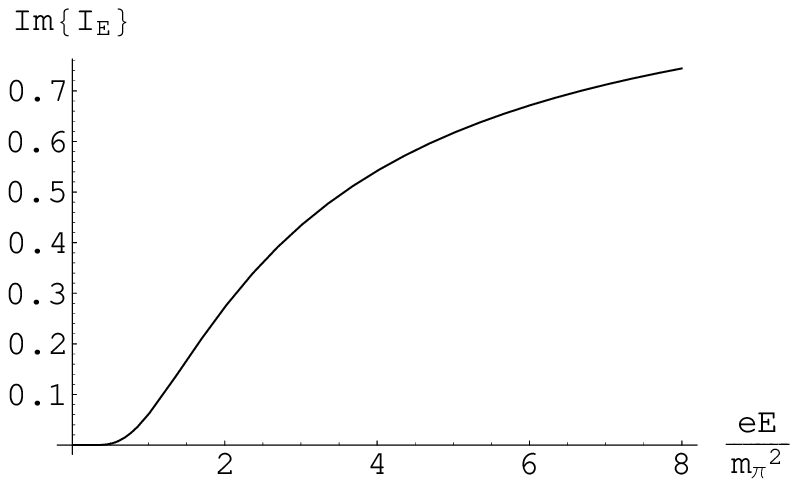}}
\\\subfigure[$Im\{I_{E}\}/Re\{I_{E}\}$ vs $eE/m_\pi^2$]{
  \includegraphics[width=0.99\linewidth]{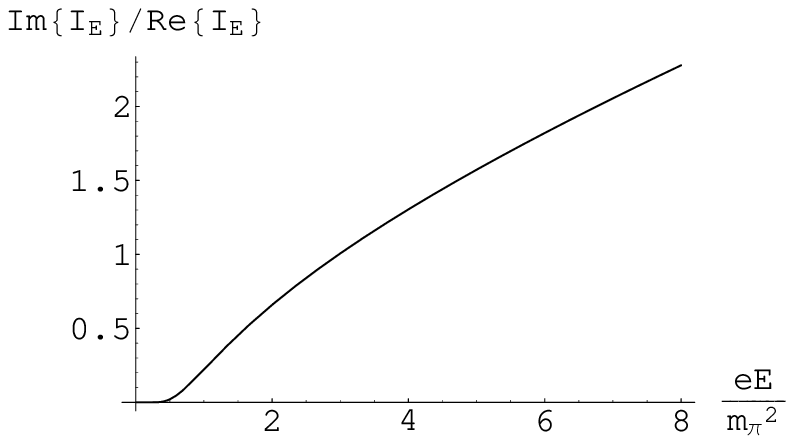}}
\caption{\label{fig:Epartplots}Real and imaginary parts of $I_E$
defined in Eq.~(\ref{eqn:exactEfromH}) are given in subfigures (a)
and (b). Subfigure (c) gives their ratio. }
\end{figure}

We turn now to the evaluation of the shift from the electric field
by analytically continuing the  closed form for $I_H$ given in
(\ref{eqn:exactH}) to find a closed form for $I_E$. Unfortunately,
the analytic structure of (\ref{eqn:exactH}) is rather
complicated; to obtain the correct analytic continuation one must
choose the appropriate branch. Making the standard choice yields
the result for the real and imaginary parts,
\begin{equation}
\begin{split}
\label{eqn:exactEfromH}
{\cal I}(I_E)  = &\frac{1}{\log 2}\log(1+e^{- \pi y})
\\ {\cal R} (I_E)  = &\frac{1}{\log 2}\Bigg\{y \log\left(\frac{y}{2}\right) - y + Cy+2\tan^{-1}\! y
\\&+ 2 \sum_{n=1}^{\infty} \left[\tan^{-1}\!\left(\frac{y}{2n+1}\right) - \frac{y}{2n}\right] \Bigg\}
\end{split}
\end{equation}
where C is Euler's constant. We use these results to find the
imaginary and real parts of $I_E$, as well as the ratio of the two
parts in Fig.~\ref{fig:Epartplots}. Note that when $e E/m_\pi^2$
is smaller than unity, the imaginary part is very small, and it is
meaningful to consider a shift in the chiral condensate due to a
``constant'' electric field.

As noted, the analytic structure of $I_E$ is complex and accordingly
it is not obvious that the choice of the branch yielding the results
in (\ref{eqn:exactEfromH}) and Fig.~\ref{fig:Epartplots} is done
correctly. In order to ensure that we have made the correct analytic
continuation, we evaluate the integral numerically to allow a direct
comparison.

Summing the residues from each of these poles we find that this
result is analytically equivalent to the imaginary part of $I_E$
in Eq.~(\ref{eqn:exactEfromH}).  This confirms our choice of
branches in the analytic continuation,
Eq.~(\ref{eqn:exactEfromH}), and shows why the imaginary part,
associated with an instability resulting from pair creation, is
unimportant for small field strengths. Expanding out the log, it
is easy to see that, as expected, the imaginary contribution is
exponentially suppressed when $eE \ll m_\pi^2$.

It is also useful to check that the real part of $I_E$ obtained via
the principal value part of the integral in Eq.~(\ref{eqn:ElecInt})
agrees with our analytic continuation.  Unfortunately, it is quite
difficult to directly evaluate the principal value of the integral
analytically. On the other hand, the singular behavior about the
poles complicates the numerical evaluation of the principal value of
$I_E$. This is neatly circumvented by changing the structure of the
integrand to:
\begin{equation}
\begin{split}
P(I_E)\! =\! &\int_{0}^{\infty} dz \Bigg\{ \frac{ e^{-y z}}{z^2} \left[
\frac{z}{\sin(z)+i\epsilon} - 1 \right]
\\&-\!\sum_{n=1}^{\infty}\frac{(-1)^{n\!-\!1}}{n \pi}e^{-n \pi y}
\!\left(\frac{1}{z \!-\! n \pi\! +\!i\epsilon} \!-\! \frac{1}{z \!+ \!n \pi\!+\!i\epsilon}\right) \Bigg\}
 \; .
\end{split}
\end{equation}

The added terms in the integrand are explicitly chosen to both have
vanishing principal values and have simple poles which exactly
cancel the poles that naturally occur in (\ref{eqn:ElecInt}).  In
this manner, the singular behavior about the poles of $1/\sin(z)$ is
eliminated from the integrand, facilitating quick and relatively
accurate numerical analysis--while simultaneously leaving the
principal value completely unchanged. The numerical results are
found to agree with the numerical values given by the formally
derived result (\ref{eqn:exactEfromH}), plotted in
Fig.~\ref{fig:Epartplots}, to extremely high precision.

To summarize, the shift in the chiral condensate at one loop in
chiral perturbation theory is given in Eqs.~(\ref{eqn:ElecInt})
and (\ref{eqn:exactEfromH}). The imaginary and real parts of the
shift are found to exactly match the pole sum and numerically
evaluated principal value, proving that the correct branch was
selected in the analytic continuation. Provided that $e E \ll
m_\pi^2$, the result is meaningful in the sense that the imaginary
part is very small and hence the instability of the vacuum due to
pair creation is only important over long times.

\section{General case: $\vec{E} \cdot \vec{H} \neq 0$}

The shifts in the chiral condensate at one loop in $\chi$PT due to a
pure uniform electric field or a pure constant magnetic field are
analytically tractable, and have closed form solutions. This is not
true in the case where both $E$ and $H$ fields are present.  If
$\vec{E}$ and $\vec{H}$ are orthogonal, one can always boost to a
frame in which the field is purely electric or magnetic, use the fact
that $\Sigma$ is a Lorentz scalar, and exploit the previous exact
results. However, where the Lorentz invariant $ \vec{E} \cdot \vec{H}$
is non-vanishing, one cannot exploit this trick and a new calculation
is needed.  In this regime, while an integral expression can be found
for $\Sigma$ at one-loop order in $\chi$PT, the integral derived from
Eq.~(\ref{eqn:SchwingerLeffBoson}) appears to be intractable
analytically.  However, the shifts may be evaluated numerically
through an extension of the approach to the cases where $E$ and $H$
are orthogonal.

As noted previously, $\Sigma$ is a Lorentz invariant and because it
is shifted by external electromagnetic fields, $\Delta \Sigma$ must
be a function of the two Lorentz scalars ${\cal F}=
\frac{H^2-E^2}{2}$ and ${\cal G }=\vec{E} \cdot \vec{H}$. Rather
than using these variables directly, it is more convenient to use
the covariant variables $f$ and $\phi$ defined according to
\begin{equation}
{\cal F}= \frac{f^2 \cos (2 \phi) }{2}  \; \; \; {\cal G}= \frac{f^2
\sin (2 \phi) }{2}
\end{equation}
with $\pi/2 \ge \phi \ge -\pi/2$.  We can reduce this domain to $0
\le \phi \le \pi/2$ by making the observation that our result is
derived from a parity-invariant effective theory for QCD; since
$E$, and therefore $\phi$, are parity-odd, our result must be an
even function of $\phi$.

One can always boost to a frame in which $\vec{E} \| \vec{H}$; in
such a frame the magnitudes $E$ and $H$ fully specify ${\cal F}$ and
${\cal G}$ up to an irrelevant sign, and have a very simple relation
to $f$ and $\phi$
\begin{equation}
H= f \cos (\phi )  \; \;  E= f \sin (\phi) \;.
\end{equation}
In such a frame, $f$ is proportional the energy density, while
$\phi$ dials the system from pure electric field aligned (or
anti-aligned) with any infinitesimal magnetic field ($\phi=\pi/2$)
to pure magnetic field $\phi=0$.

Using the Schwinger proper time expression for the effective
Lagrangian of Eq.~(\ref{eqn:SchwingerLeffBoson})for the case of
non-orthogonal $E$ and $H$ fields and differentiating with respect
to $m_\pi^2$ to obtain the chiral condensate yields
\begin{equation}\label{eq:gen}
\Delta \Sigma (f, \phi) = \frac{ef\Sigma(0)\log 2}{16 \pi^2 F_{\pi}^2}
I_{EH}(f, \phi),
\end{equation}
where $I_{EH}(f, \phi)$ is defined to be
\begin{widetext}
\begin{eqnarray}
\label{eqn:EHfull} I_{EH}(f, \phi) & = &\frac{1}{\log 2} \int_{0}^{\infty}
\frac{du}{u^2} e^{-(m_\pi^2/e f) u} \left( \frac{u^2 \sin (2 \phi)}{2
\sin\left(u
\sin(\phi)\right)\sinh\left(u\cos(\phi)\right)+i\epsilon} - 1
\right)
\\{\cal R}(I_{EH}(f, \phi)) & = &I_{EH} - \frac{1}{\log 2}\sum_{n=1}^{\infty}\frac{\cos(\phi)(-1)^{n}e^{-n \pi m_\pi^2/ (e f \sin(\phi))}}{\sinh(n \pi /\tan(\phi))} \int_{0}^{\infty}\left(\frac{du}{u - \frac{n \pi}{\sin(\phi)}+i\epsilon}
- \frac{du}{u + \frac{n \pi}{\sin(\phi)}+i\epsilon}\right)\label{eqn:EHre}
\\{\cal I}(I_{EH}(f, \phi)) & = & \frac{1}{\log 2}\sum_{n=1}^{\infty}\frac{\pi \cos(\phi)(-1)^{n}e^{-n \pi m_\pi^2/ (e f \sin(\phi))}}{\sinh(n \pi /\tan(\phi))}.
\label{eqn:EHim}
\end{eqnarray}
\end{widetext}
We find the integral in Eq.~(\ref{eqn:EHfull}) to be intractable
analytically and thus we evaluate it numerically. Using the same
$i\epsilon$ convention as with $I_E$, the integral can be divided
into a principal value part and a contribution from the residue of
the poles.  To evaluate the principal value part, we use the same
prescription for removing the poles as in Sec.~\ref{sec:Efield}.
The principal value (real part) is given in Eq.~(\ref{eqn:EHre})
and the sum of the residues (imaginary part) is given in
Eq.~(\ref{eqn:EHim}).

\begin{figure}[tbp]
\subfigure[${\cal R}(I_{EH})$ vs $e f/m_\pi^2$]{
  \includegraphics[width=.99\linewidth]{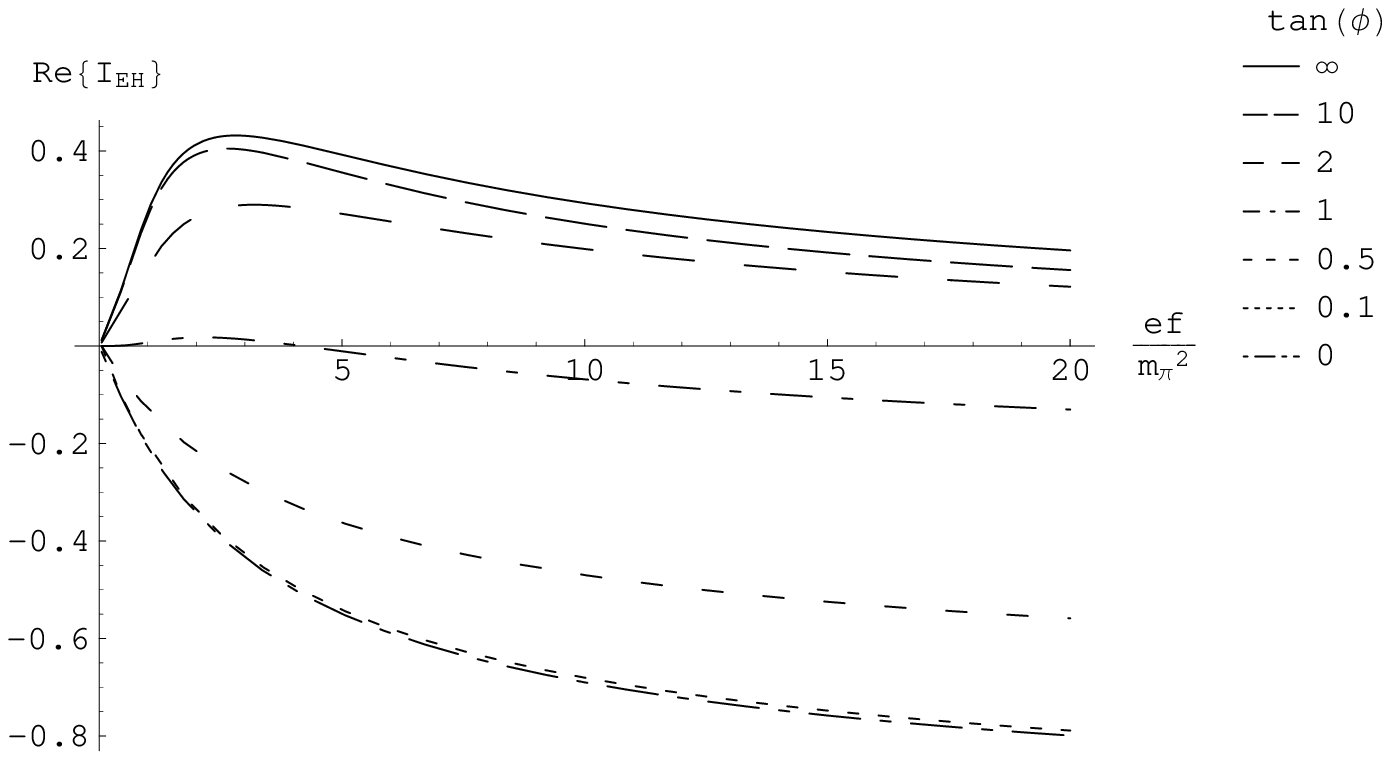}}
\\\subfigure[${\cal R}(I_{EH})$ vs $m_\pi^2/e f$]{
  \includegraphics[width=.99\linewidth]{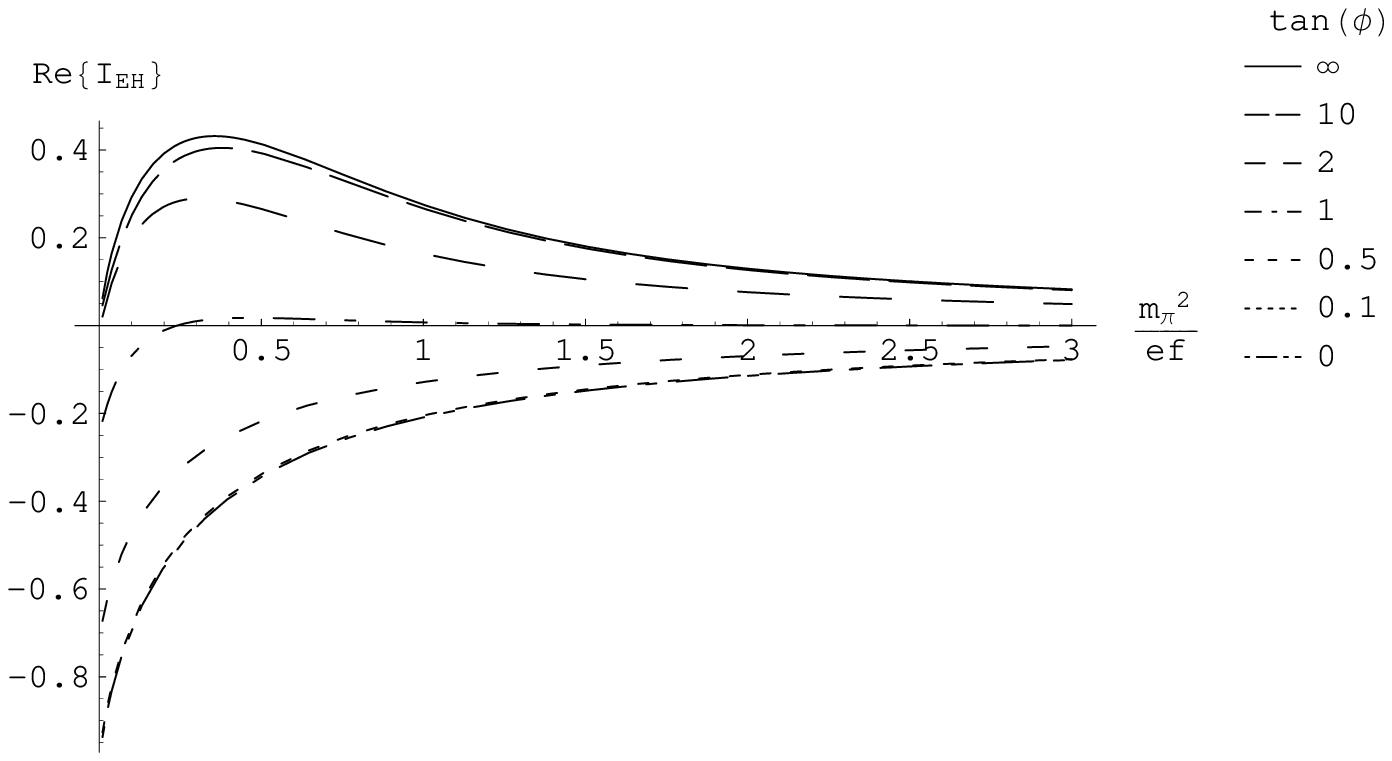}}
\caption{\label{fig:EHprincipal}Principal value of the shift in
the condensate plotted against the Lorentz invariants $\tan(\phi)$
and $e f$ as defined in the text.}
\end{figure}

\begin{figure}[tbp]
\subfigure[${\cal I}(I_{EH})$ vs $e f/m_\pi^2$]{
  \includegraphics[width=0.99\linewidth]{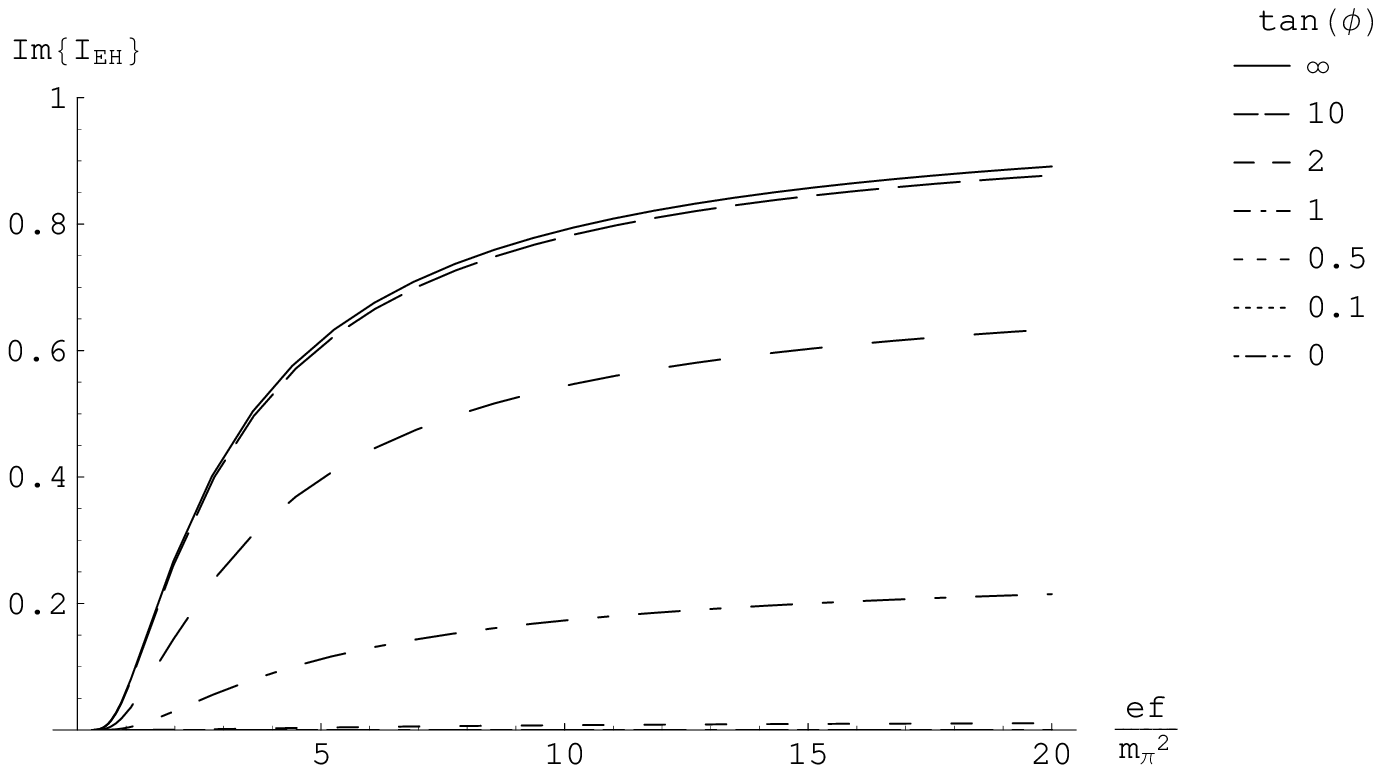}}
\\\subfigure[${\cal I}(I_{EH})$ vs $m_\pi^2/e f$]{
  \includegraphics[width=0.99\linewidth]{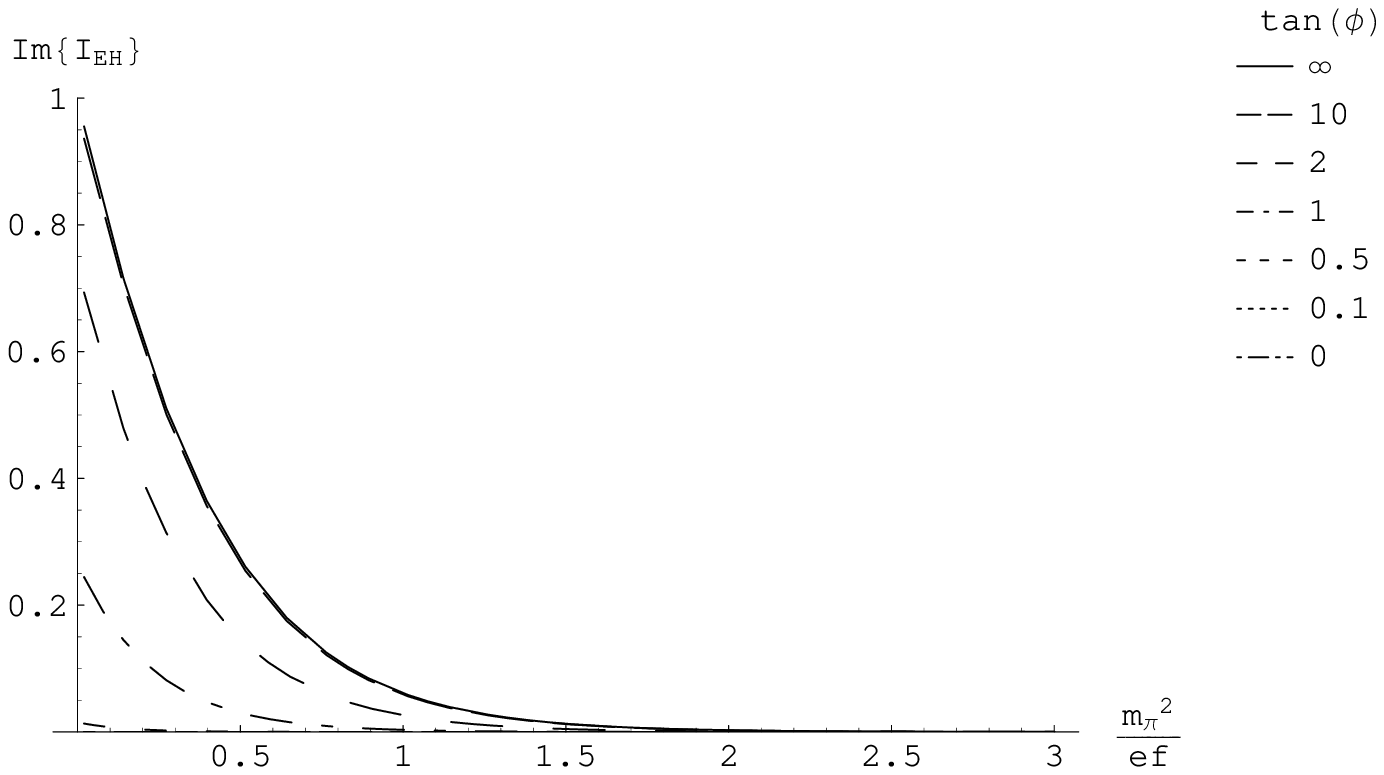}}
\caption{\label{fig:EHpoles} Sum of the residues of the shift in
the condensate plotted against the Lorentz invariants $\tan(\phi)$
and $e f$ as defined in the text.}
\end{figure}

In Fig.~\ref{fig:EHprincipal}, we plot the principal value of this
integral as a function of $\tan(\phi)$ and $e f$. We see that ${\cal
R}(I_{EH}(f, \phi))$ approaches ${\cal R}(I_E)$ at one extreme and
$I_H$ at the other extreme, as expected. Fig.~\ref{fig:EHpoles}
shows the imaginary part arising from the pole structure for various
values of $E$ and $H$. We note a smooth variation from no imaginary
part for a pure magnetic field to the maximal imaginary part for a pure
electric field.

We also see that the imaginary part (and therefore the pair creation
from the electric field) is suppressed at high mass in all cases.
This means that far from the $m_\pi=0$ limit, the pair creation
mechanism will not play a role and the constant field will be nearly
stable over relevant time scales.

For $m_\pi^2/e f$ large enough so that the imaginary part is
negligible, we also see that $I_{EH}(f, \phi) \rightarrow -I_{EH}(f,
\phi)$ under $\phi \rightarrow \pi/2 - \phi$, which corresponds to
switching the electric and magnetic fields. Physically, this means
that at high $m_\pi^2/e f$ where the imaginary part is suppressed,
the effect of switching the electric and magnetic fields on the
shift in the condensate introduces an overall negative sign. Thus,
we see the general effect that while the magnetic field acts to
increase chiral symmetry breaking, the electric field suppresses it.

\begin{figure}[tbp]
  \includegraphics[width=0.99\linewidth]{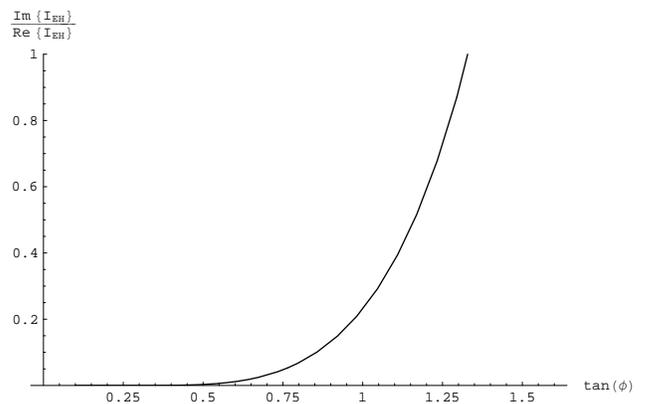}
\caption{\label{fig:EHrat}Ratio of the imaginary pair-creating piece
versus the principal value, plotted as a function of $\tan(\phi)$
for $m_\pi^2/e f = 0$.}
\end{figure}

In Fig.~\ref{fig:EHrat}, we plot the ratio of the imaginary to real
parts as a function of $\tan(\phi)$, which, again, corresponds to
$E/H$ in the limit of parallel fields, now in the opposite extreme
where $m_\pi^2/e f = 0$. The imaginary part grows exponentially with
$\tan(\phi)$. It is negligible for $\tan(\phi) \lesssim 0.5$, and
becomes significant by $\tan(\phi) \sim 1$.

\section{Concluding Remarks}

The general expression for the shift in the chiral condensate in
QCD due to electric and magnetic fields is given in
Eqs.~(\ref{eq:gen}) and (\ref{eqn:EHfull}). The expression is
valid mathematically up to power corrections in
$m_\pi^2/\Lambda^2$ and $e f/\Lambda^2$ and hence may be regarded
as a ``low energy theorem'' in the same sense used by Shushpanov
and Smilga\cite{Smilga}. This expression may appear to be
dynamics-dependent, in as much as it depends on the specific
Lagrangian of $\chi$PT, but as the $\chi$PT Lagrangian is the only
possible Lagrangian consistent with the symmetries of QCD possible
at low energies, this is not the case.

A non-trivial uniform electric field, which occurs when ${\cal G} \neq
0$ or ${\cal F} < 0$, causes poles to appear in the expression for the
shift in the chiral condensate. As a result, when Schwinger's boundary
conditions are imposed, the shift acquires a non-zero imaginary
component, which corresponds to an instability in the vacuum. On a
physical level, the instability of the vacuum in a constant electric
field restricts the applicability of the result to cases where the
instability occurs over a relatively long time scale. This essentially
restricts one to the range $e E \ll m_\pi^2$, in the frame where $E \|
H$, for which the imaginary part is exponentially suppressed. In
general, the integral in Eq.~(\ref{eqn:EHfull}) needs to be treated
numerically. However, for the special cases of pure electric and pure
magnetic fields, the integral may be evaluated analytically, leading
to the simple expressions in Eqs.~(\ref{eqn:exactH}) and
(\ref{eqn:exactEfromH}). These results generalize previous work both
via the inclusion of an electric field and through the inclusion of a
non-zero pion mass. The need to include a non-zero pion mass is
critical since the weak field and zero $m_\pi$ limits are non
uniform. As a result, the behavior at $e f \sim m_\pi^2$, the
principal regime of relevance, in our chiral expansion is typically
quite different from the behavior at fixed $f$ but zero $m_\pi$.

Clearly, since our expression is just the leading-order term in
the chiral expansion it has a limited range of validity in field
strength. To improve the accuracy of the description at somewhat
larger field strengths, it is natural to work to higher order in
the expansion. Such a calculation has been done for purely
magnetic fields at $m_\pi=0$ \cite{Shushpanov}; it would certainly
be of interest to extend this to the general case of electric and
magnetic fields to all orders in the ratio of $m_\pi^2/(e f)$.

Before concluding, it is worth noting that the NJL model
calculations \cite{Klevansky} find a first order
response in a magnetic field and zero $m_\pi$ to be $\propto
(eH)^2/\Sigma(0)^4$, which is not consistent with the $\chi$PT
result. One might expect that the $\chi$PT result, based on QCD in
a model independent way, is obviously more accurate then the NJL
result which is clearly model dependent. However, the NJL result
is of some interest because the nature of the calculations region
of validity differs from that of $\chi$PT. It is derived in the
Hartree-Fock approximation, justified by the $1/N_c$
expansion\cite{tHooft,Witten}. In contrast, $\chi$PT is an
expansion in small momentum and quark masses, but its $N_c$
dependence is less obvious. The low-energy constants (LECs) which
multiply each term in the expansion are independent of momentum,
but can have non-trivial $N_c$ dependence. Thus, though higher
order terms in the expansion may appear to drop with $N_c$, due to
inverse powers of $F_\pi$, this is not actually always the
case\cite{GasserLeutwyler}. One question is whether or not the
large $N_c$ and chiral limits commute for $\Sigma(H)$. It would
hardly be surprising if the limits do not commute as there are
many well-known examples in QCD of non-commuting limits
\cite{limits}. In any case, the NJL results are useful as a hint
that the region of validity for $\chi$PT might not be as clean cut
as might be hoped. Higher-order terms in the $\chi$PT expansion
which are not $1/N_c$ suppressed can cause the expansion to
converge more slowly than would otherwise be expected.

Thus, another motivation for working to higher order is provided
by the non-trivial $N_c$ dependence of the LECs noted above. The
calculations done here came entirely from a pion loop, and meson
loops are generically suppressed by factors of $1/N_c$. In
contrast, tree diagrams in the chiral Lagrangian can yield leading
order results in the $1/N_c$ expansion \cite{Bijnens}.  As noted
in Sec.~\ref{sec:mag}, it is typical for tree graphs from the
${\cal O}(p^4)$ chiral Lagrangian to contribute at the same order
in the chiral expansion as one-pion-loop graphs.  However, for the
shift in the chiral condensate, there are no terms in the ${\cal
O}(p^4)$ Lagrangian which contribute at tree level.  This is not
the case at the next order in $\chi$PT where tree graphs in ${\cal
O}(p^6)$ Lagrangian do contribute \cite{Shushpanov}; it is easy to
see these contributions come in at leading order in the $1/N_c$
expansion. An analysis of the actual impact of these leading-order
$N_c$ terms may shed light on whether or not the $N_c$ expansion
can indicate a slower convergence of $\chi$PT.

In conclusion, we have numerically examined the shift in the QCD
chiral condensate due to an electromagnetic field in the framework of
Chiral Perturbation Theory. We find that the low energy theorem in
(\ref{eqn:lowenergy}) is indeed accurate at large $eH/m_\pi^2$.
However, a field which is large enough for this limit to apply is
large enough that an expansion in $eH/F_\pi^2$ is no longer
valid. These results continue to hold in the presence of a small
electric field.  We are thus able to demonstrate, in a model
independent way, that uniform electric fields tend to suppress
S$\chi$SB. However, when the magnitude of the electric field becomes
comparable to that of the magnetic field (in the frame where they are
both parallel), the calculations begin to break down due to the
imaginary shift which corresponds to local instabilities and charged
pair creation.

This work is supported by the U. S. Department of Energy under
grant number DE-FG02-93ER-40762.

\clearpage

\end{document}